\documentclass[]{spie}  %>>> use for US letter paper
\usepackage{amsmath,amsfonts,amssymb}
\usepackage{graphicx}
\usepackage[colorlinks=true, allcolors=blue]{hyperref}

\title{Automated Project Completion Forecasting}

\author[a]{Alexandra J. Tetarenko}
\author[a]{Harriet Parsons}
\author[a]{Sarah Graves}
\author[a]{Jessica Dempsey}
\affil[a]{East Asian Observatory, 660 N. A'oh\={o}k\={u} Place, University Park, Hilo, Hawaii 96720, USA}

\authorinfo{Further author information: Send correspondence to A.J. Tetarenko, E-mail: a.tetarenko@eaobservatory.org.}

\begin{document} 
\maketitle

\begin{abstract}
In the age of Large Programs and Big Data a key component in project planning for ground-based astronomical observatories is understanding how to balance users demands and telescope capabilities. In particular, future planning for operations requires us to asses the impact of a complex set of parameters, such as right ascension, instrument, and sky condition pressures over coming semesters. Increased understanding of these parameters can provide: improved scientific output, better management of user expectations, more accurate advertised/allocated time under a Call for Proposals, and improved scheduling for instrumental commissioning and engineering work. We present  ongoing efforts by staff at the James Clerk Maxwell Telescope (JCMT) to build a tool to provide automated completion forecasting of Large Programs undertaken at this telescope, which make up 50\% of the observing time available at the JCMT.
\end{abstract}


\section{INTRODUCTION}
\label{sec:intro}
The James Clerk Maxwell Telescope (JCMT) is a 15 meter single-dish sub-millimetre telescope located on the slopes of Maunakea in Hawaii, at a latitude of 19.8 degrees and elevation of 4092 meters. The JCMT has a suite of instruments for continuum observations (SCUBA-2\cite{holl13,demp13}, operating at 450/850 microns), polarization observations (POL-2\cite{free18}, operating at 450/850 microns), and heterodyne observations (HARP\cite{buck09}, operating between 325--375\,GHz, and N\=amakanui, with the newly commissioned $`$\=U$`$\=u receiver operating between 221--265\,GHz). With these instruments, the JCMT can observe a vast range of different science targets, including solar system objects\cite{greaves20}, the star formation process\cite{mairs19}, evolved stars\cite{dharm20}, black holes and their outflows\cite{tetarenkoa17,eht}, gamma-ray bursts\cite{smith19}, and other energetic transients\cite{smith18}.

Many open questions in astrophysics can only be answered after years of data collection at a telescope. Therefore, we have seen a push in many different astronomical fields towards obtaining such large data sets. In 2007, the JCMT responded to these user needs by implementing the JCMT Legacy Surveys\footnote{\url{https://www.eaobservatory.org/jcmt/science/legacy-survey/}}. These surveys\cite{Chrysostomou2012} were a set of seven observing programs that ran for several years and produced over 119 publications \cite{hatch13,vdw11,war10,thom07,geach13,hol17}. In 2015, following the East Asian Observatory's (EAO) takeover of operations, JCMT began a new set of Large Programs. These new programs acted as a continuation of the highly successful Legacy Surveys, and were to additionally provide a training ground for the new EAO user community. Currently, these Large Programs account for 50\% of observing time available to the JCMT community, and during the past 5 years have produced $\sim$10--25\% of the total publications using JCMT data.

JCMT operates on a two-semester per year observing schedule, where the A semester runs from February 1 -- July 31, and the B semester runs from August 1 -- January 31. In each semester, in addition to the Large Programs (LAP), the JCMT offers three more types of observing programs; Principle Investigator (PI; competitive, merit-based time available to all astronomers in JCMT's funding regions\footnote{JCMT's funding regions include: Canada, China, Indonesia, Japan, Korea, Malaysia, Taiwan, Thailand, UK, and Vietnam.}), directors discretionary time (DDT), and University of Hawaii time (UH; time reserved for faculty/students at the University of Hawaii). JCMT follows a Flexible Queue structure (sometimes referred to as ``dynamic scheduling"\cite{robson2002}) to balance the number of nights for each type of observing program per semester, whereby each type of program is allocated a block of nights in a rotating sequence in the telescope schedule.

To maximize the scientific output of JCMT, we need to ensure that the telescope is operating as efficiently as possible\cite{demp2018}. The observatory already has in place a set of tools that automates the proposal--execution--data reduction--project completion process for single semester PI observing programs. However, for Large Programs,
staff at the JCMT have been manually tracking the progression of these multi-semester projects, relying on many simplifying assumptions (e.g., approximate right ascension ranges of science targets, historical weather data) to make loose predictions about completion rates and available observing time at the start of each semester. This task was already an extensive job in the Legacy Survey era, and has quickly become not feasible anymore with up to 19 Large Programs active in the queue (see Table~\ref{table:lp2019}). 
As such, we need new methods to accurately simulate JCMT observing, and ultimately estimate the amount of observing time we can offer to our community each semester. 
In this paper, we introduce the {\sl JCMT Automated Project Completion Forecaster}, a \textsc{python} program that simulates Large Program observing in future semesters at the JCMT.

\section{Large Program Observing at JCMT}
\label{sec:lp}

Large Programs are observing projects that have been chosen by the JCMT board and the JCMT Time Allocation Committee for their high science impact. These programs require hundreds of hours to reach their scientific goals, span multiple semesters, and are open to the wider JCMT community to join and contribute towards. The overarching goals of the Large Programs are to make significant contributions to some of the most pressing problems in modern astrophysics, as well as engage and train the ever growing user community on the capabilities of the suite of instruments offered at JCMT. Since their inauguration, the JCMT Large Program scheme has awarded over 6,000 hours of telescope time to 18 \textit{unique} science programs. %split across 29 different observing projects. 

\subsection{Program Types}
\label{sec:progtype}
The current selection of JCMT Large Programs target a range of scientific topics, including evolved stars, star formation within the Milky Way, nearby galaxy studies, cosmological studies of sub-millimetre galaxies at high redshifts, and black hole studies\footnote{\url{https://www.eaobservatory.org/jcmt/science/large-programs/}}. These programs utilize all of the available JCMT instruments, and include several different types of observations (continuum, spectral line), with different observing constraints (fixed cadence [Transient], low elevation observations [NESS]). Additionally, PITCH-BLACK is a target of opportunity (ToO) project, which aims to observe transient events, and thus will only be observed when specific target source trigger conditions are met. Table~\ref{table:lp2019} summarizes the current suite of approved JCMT Large Programs.

\renewcommand\tabcolsep{2pt}
 \begin{table}[t!]
\caption{JCMT Large Programs active in the telescope queue}\quad
\centering
\begin{tabular}{lp{3.7cm} cccc}
 \hline\hline
 {\bf Program Name -- Code$^a$}&{\bf Instrument(s) }&{\bf Allocated}&{\bf Science}&{\bf Weather}\\
  {\bf }&{\bf }&{\bf Time (Hrs)$^b$}&{\bf Area}&{\bf Grade}$^c$\\[0.15cm]
  \hline
 PITCH-BLACK -- M20AL008& SCUBA-2/POL-2 & 256 & black holes & 1--3\\[0.05cm]
 Transient -- M20AL007&SCUBA-2&273 [200]&star formation&1--3\\[0.05cm]
 HASHTAG -- M17BL005&SCUBA-2/HARP&[276]&nearby galaxies&2-3\\[0.05cm]
 STUDIES -- M17BL009&SCUBA-2&[319+330]&sub-millimetre galaxies&1\\[0.05cm]
 CHIMPS -- M17BL004&HARP&[404]&star formation&1,3,4\\[0.05cm]
 MALATANG -- M20AL022$^d$&HARP&404 [390]&nearby galaxies&3\\[0.05cm]
 BISTRO -- M20AL018/M17BL011&POL-2&224 [224+224]&star formation&1--2\\[0.05cm]
 NESS -- M20AL014/M17BL002&SCUBA-2/HARP/$`$\=U$`$\=u&748+[515]&evolved stars&1--5\\[0.05cm]
 ALOHA IRDCs -- M20AL021&SCUBA-2&130&star formation&2--3\\[0.05cm]
 RAGERS -- M20AL015&SCUBA-2&168&sub-millimetre galaxies&1--2\\[0.05cm]
 DOWSING -- M20AL011&SCUBA-2&240&nearby galaxies&1--2\\[0.05cm]
 JINGLE -- M20AL013/M16AL005&SCUBA-2/$`$\=U$`$\=u&445+[780]&nearby galaxies&2--5\\[0.05cm]
 NEP -- M20AL005&SCUBA-2&200+[200]&sub-millimetre galaxies&3\\[0.05cm]
 SPACE -- M20AL006&POL-2/$`$\=U$`$\=u&454&star formation&2,4\\[0.05cm]
 AWESOME -- M20AL020&SCUBA-2&100&sub-millimetre galaxies&1\\[0.05cm]
 S2LXS -- M20AL026&SCUBA-2&189&sub-millimetre galaxies&2\\
 \hline
\end{tabular}\\
\begin{flushleft}
$^a$ If more than one program code is given, then a previous iteration of the program is still active in the Large Program queue.\\
$^b$ Square brackets indicate time allocated to the specified Large Program prior to the latest Large Program Call for Proposals in February 2020. For example, the BISTRO program is currently on its third iteration, having received time in two earlier Large Program Calls for Proposals, while the HASHTAG, STUDIES, and CHIMPS programs received time in a previous Large Program Call for Proposals and have not completed observing yet.\\
$^c$ JCMT weather grades are defined in \S\ref{sec:wb}.\\
$^d$ The MALATANG program is not currently being observed, as it is awaiting JCMT Board approval to begin observing its 2020 allocation of time.\\
\end{flushleft}
\label{table:lp2019}
\end{table}
\renewcommand\tabcolsep{6pt}

\subsection{Weather Grades}
\label{sec:wb}
At sub-millimetre wavelengths, the main weather-based constraint that defines if we can observe a given project is the atmospheric opacity due to water vapour. As the opacity increases, more time is required on sky to reach the same sensitivity at a given wavelength. At JCMT, we define several weather grades based on the measured amount of precipitable water vapour (PWV) along our line-of-sight through the atmosphere. To measure PWV, we use an in-cabin line-of-sight 183 GHz water vapor radiometer (WVM), allowing us to track the atmospheric conditions with high time-resolution. The JCMT weather grades are defined as follows: Grade 1 (PWV$<$0.83 mm), Grade 2 (0.83 mm$<$PWV$<$1.58 mm), Grade 3 (1.58 mm$<$PWV$<$2.58 mm), Grade 4 (2.58 mm$<$PWV$<$4.58 mm), and Grade 5 (PWV$>$4.58 mm). In general, summer months on Maunakea (May--Nov; 5\% of the time spent in Grade 1) tend to be wetter than winter months (Dec--April; 25\% of the time spent in Grade 1), and up to 15\% of the observing time during a given semester can be lost entirely to bad weather, where nothing can be observed.

\subsection{Calibration and Overheads}
\label{sec:calover}
The total observing time for a given night at JCMT will be made up of: (1) science targets, (2) calibration observations (typically amounting to $\sim$25\% of the science target observing time), (3) engineering and commissioning projects (this results in blocked out nights where no science targets are observed), and (4) additional overheads such as the time to slew between targets and instrumental faults ($\sim$3\% is a typical fault rate for JCMT per night). Each of these aspects must be included when simulating a real observing night at JCMT.

\subsection{Observing Nights and Dynamic Scheduling}
\label{sec:flex}

A typical observing night at JCMT includes a 13-hour shift from 0330 to 1630 UT (0530 to 0630 HST). However, in the case where the weather conditions remain stable (Grade 3 or better) into the early morning hours, we will perform extended observing (EO)\cite{Walther2014}, whereby we continue observing into the morning for up to 4 more hours (until 2030 UT or 1030 HST). Within the EO hours, we have an additional sun separation observing constraint (minimum sun separation of $45$ degrees).

To maximize productivity at JCMT, we have adopted dynamic scheduling to determine which project should be observed at any given time of the night. In this plan, the telescope operator schedules programs by priority (as defined by rankings from the Time Allocation Committee) for targets belonging to the specified nightly queue (PI, LAP, UH, DDT), that are observable at JCMT in the current weather grade. This task is performed by the telescope operator using the Query tool. The one notable exception to this plan is in the case of ToOs. If a ToO is triggered, then it immediately assumes top priority within the approved weather grade.

\begin{figure}[t!]
\centering
\includegraphics[width=0.95\textwidth]{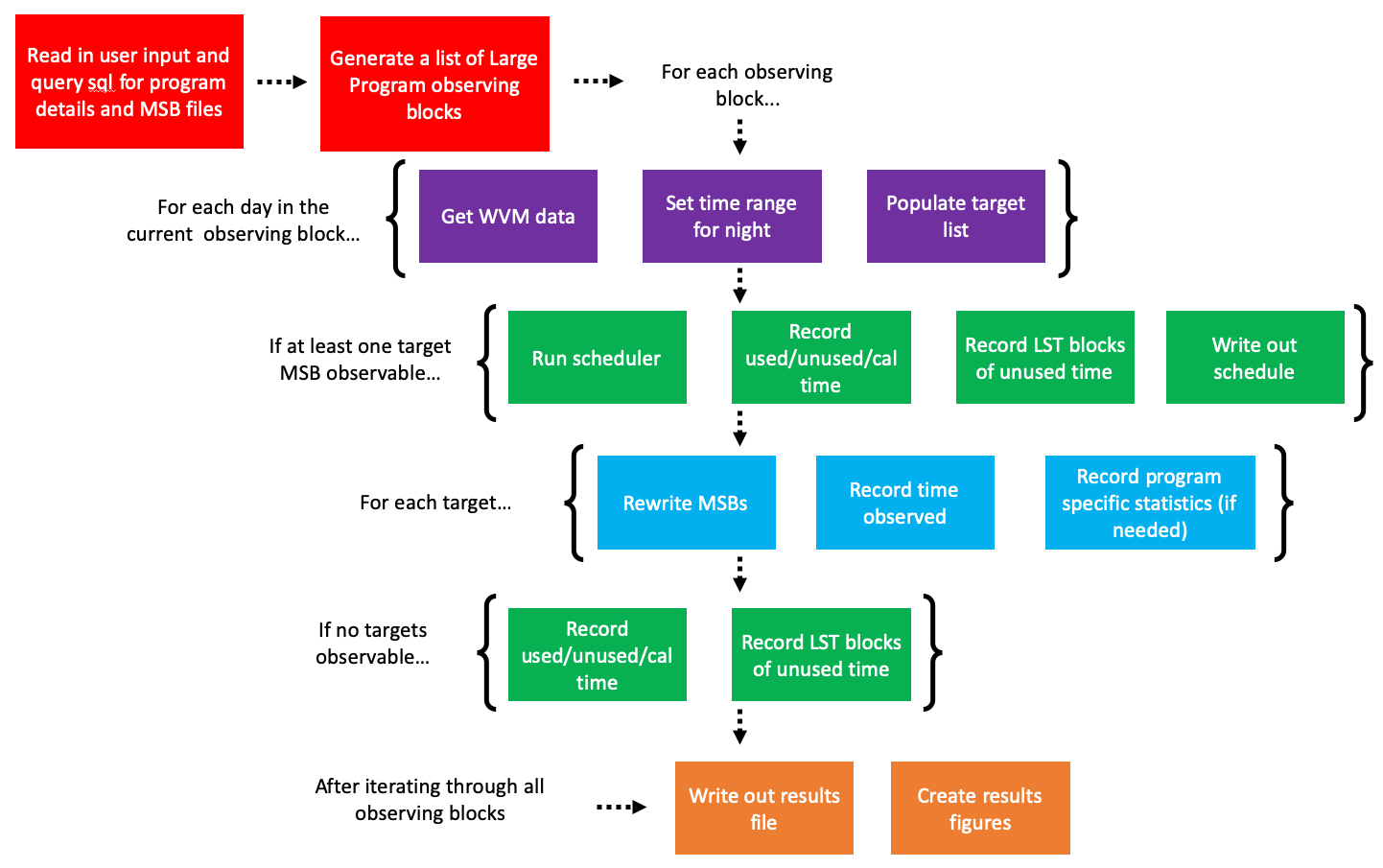}
\caption{Flow chart displaying the workflow for the {\sl JCMT Automated Project Completion Forecaster}.
}
\label{fig:workflow}
\end{figure}

\section{Automated Project Completion Forecasting Tool}
\label{sec:forecast}
%explain how the code works, include example outputs

\subsection{Overview and Workflow}
\label{sec:workflow}
The {\sl JCMT Automated Project Completion Forecaster}\footnote{\url{https://github.com/tetarenk/LP_predict}} is a \textsc{python} program that simulates JCMT Large Program observing. The script utilizes the \textsc{astroplan} package\footnote{\url{https://astroplan.readthedocs.io/en/latest/}} framework, along with user defined start/end dates for the simulation, historical WVM data to model weather patterns, and the scheduling block files produced for each JCMT Large Program (known as MSBs), which are stored in an sql database at the observatory. With these inputs, we are able to create a simulated observing schedule for each night allocated to Large Programs over the upcoming semester(s). Additional overheads due to calibrations, time to slew between targets, instrumental faults, blocked out dates for engineering/commissioning work (manually set by the user), and EO are also included in the observing schedules to more closely match a real observing night at JCMT.

We model the JCMT queue structure based on the average number of nights in each of the PI, LAP, UH, and DDT queues per semester since 2015\footnote{Following the sequence: LAP, PI, LAP, PI, LAP, PI, LAP, PI, UH, DDT, where LAP/PI are five night blocks, UH is a four night block, and DDT is one night.}.
The simulation mimics the JCMT's dynamic observing by stepping through the night sequentially, scheduling the highest priority projects that are observable. A target is only added to the target list for a night if it meets the following requirements: the target has remaining observing time, the night is not in the blackout dates for engineering/commissioning, and the weather is appropriate for the project (i.e., in the specified weather grade or better). Target priority is set as follows: (1) rank of the target's Large Program, (2) the target has the same weather grade as the night, (3) the target is allocated time for a worse weather grade.
%Prior to running the simulator, the code double checks that the total target observing time within the scheduling blocks for each project matches the total allocation, and automatically corrects any mismatches. 
The simulator automatically updates scheduling block files for all the targets observed after each simulated night, keeps track of the schedules for each night (in tabular and plot form), and records several different observing statistics, such as, how many hours are observed for each Large Program, the time remaining for each Large Program after the simulated semester(s), and unused time in each weather grade. Figure~\ref{fig:workflow} summarizes the workflow for the simulation.

\subsection{Dependencies}
\label{sec:dep}
The {\sl JCMT Automated Project Completion Forecaster} code is compatible with \textsc{python} 2 \& 3, and uses the following \textsc{python} packages: \textsc{astroplan}, \textsc{numpy}, \textsc{matplotlib}, \textsc{astropy}, \textsc{datetime}, \textsc{pandas}, and \textsc{mysql-connector-python}.

\begin{figure}[t!]
\centering
\includegraphics[width=0.85\textwidth]{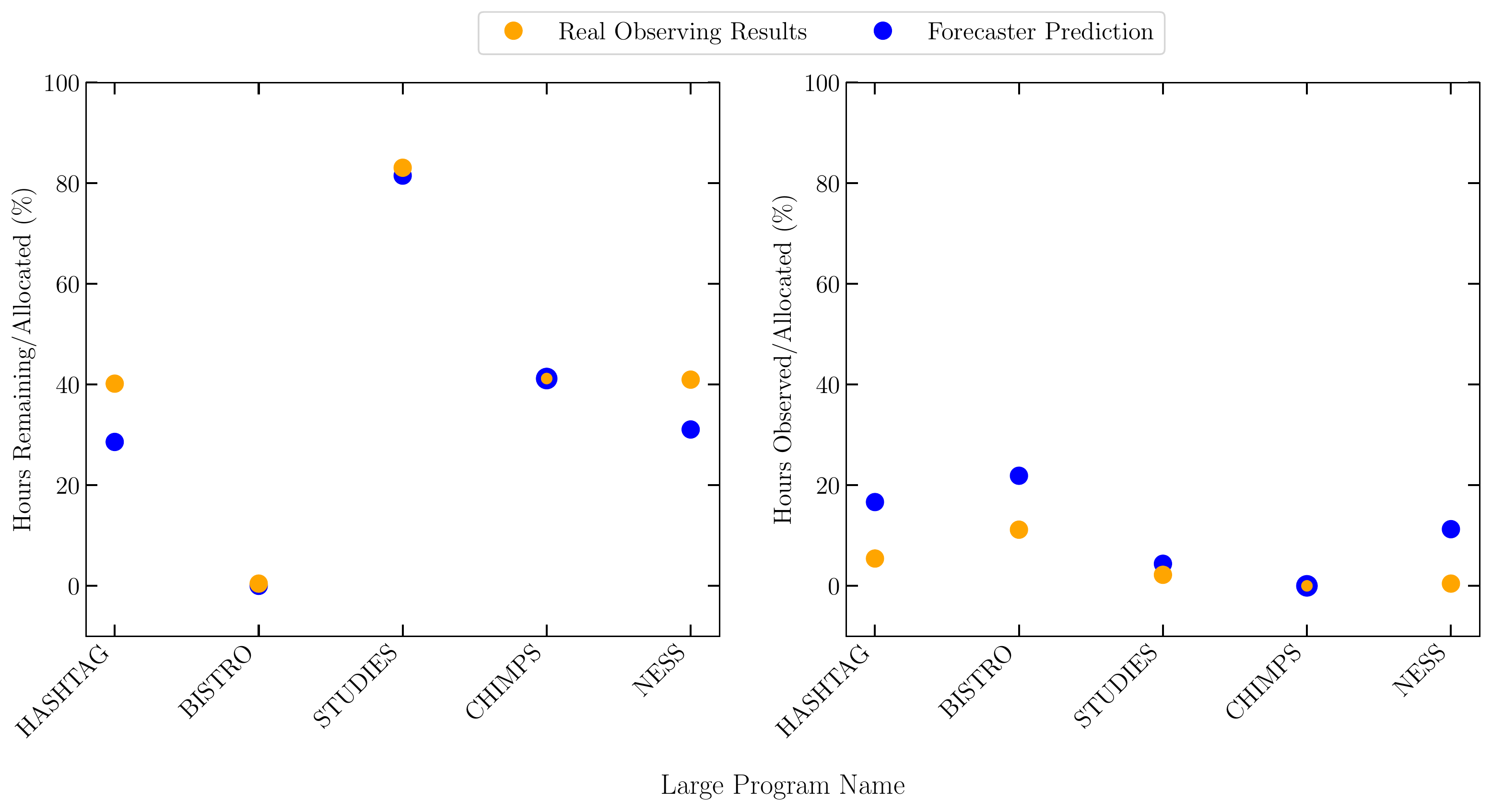}
\caption{Comparison between a
{\sl JCMT Automated Project Completion Forecaster} simulation run over a 6-month period and real JCMT observing results. {\it Left:} Hours remaining in a selection of Large Programs after the simulation/observing period. {\it Right:} Hours observed for a selection of Large Programs within the simulation/observing period.
The forecaster slightly over-predicts the time spend observing these programs. However, when comparing real world weather patterns to those in the simulation, we find that the real world weather was worse than in the simulation. This resulted in $\sim$13\% more time on sky available for Large Programs in the simulation, which could account for the over-prediction here.
While this is only a small sample of programs, over a short timescale simulation run, overall the forecaster does appear to closely match real JCMT observing results.
}
\label{fig:hrs_comp}
\end{figure}

\subsection{Testing the Forecaster}
\label{sec:test}
To test the accuracy of the 
{\sl JCMT Automated Project Completion Forecaster}, we need to compare the simulation hours of the Large Programs to real JCMT observing results, over different timescales and semesters. While this testing process is ongoing, early comparisons have shown encouraging results. For example, Figure~\ref{fig:hrs_comp} shows this comparison for a 6-month long simulation run up to February 2020 (before the new set of Large Programs, indicated by a program code beginning with ``M20" in Table~\ref{table:lp2019}, started observing). We are currently working to do similar comparisons across simulations run for longer timescales, as well as examine the breakdown of simulation hours and real world observing hours by instrument and weather grade.

\subsection{Forecasting Tool Outputs}
\label{sec:inout}

\begin{figure}[t!]
\centering
\includegraphics[width=1\textwidth]{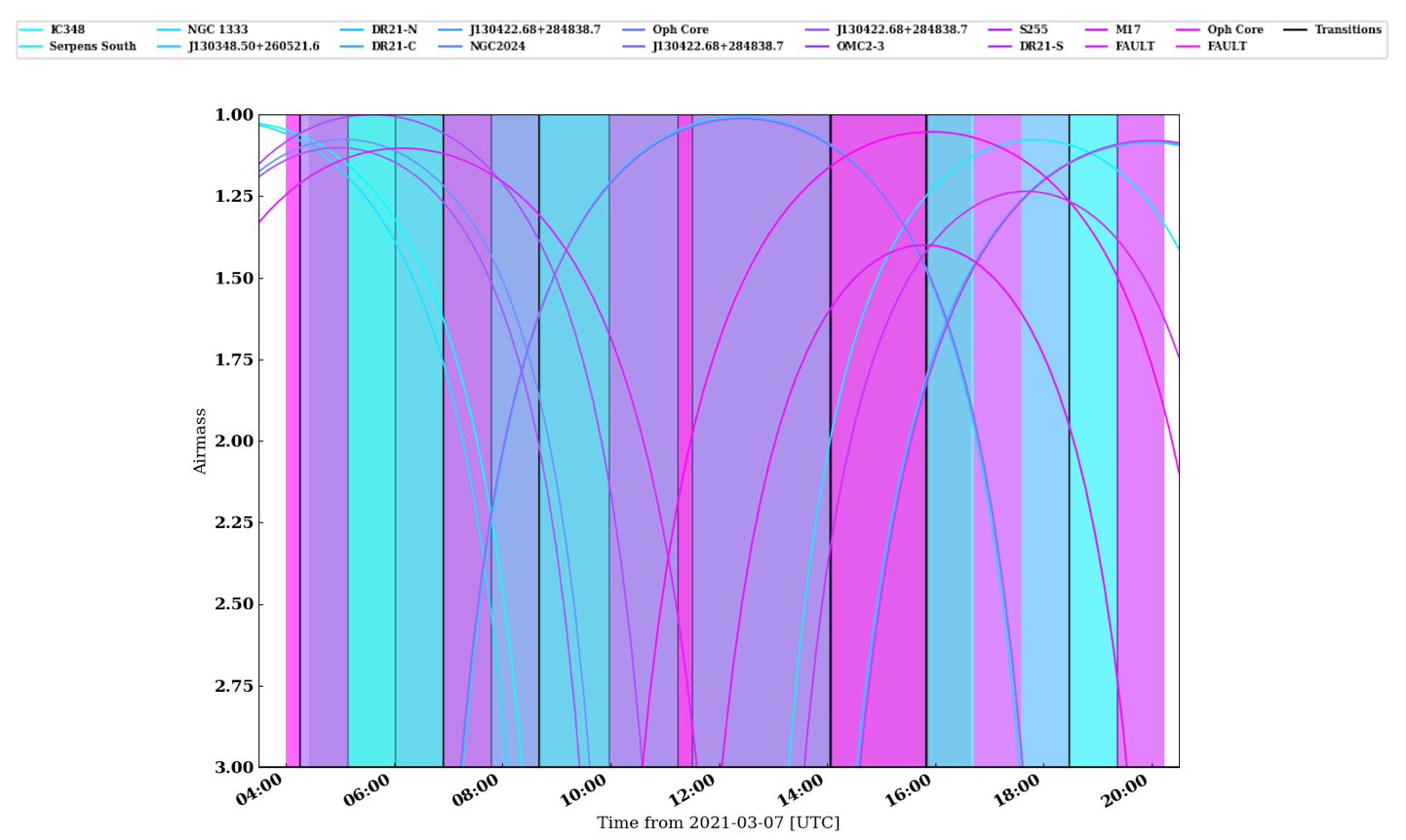}
\caption{Example schedule for a single night of simulated JCMT Large Program observing with the {\sl JCMT Automated Project Completion Forecaster}. Observing blocks (shaded regions; including science target blocks and fault blocks) and science target visibility curves (solid coloured lines) are shown, where colours represent priority (purple indicates the highest priority science targets). Solid black lines indicate transition blocks when the telescope was slewing between science targets. As this was a Grade 3 weather night in the simulation, the Extended Observing (EO) option was activated, and the simulator observed into the early morning hours. Note that the time for calibrations is included within the total time used for each science target observing block. The simulator clearly maximizes the time available for science during an JCMT observing night.
}
\label{fig:ex_sched_201917}
\end{figure}

\begin{figure}[t!]
\centering
\includegraphics[width=0.8\textwidth]{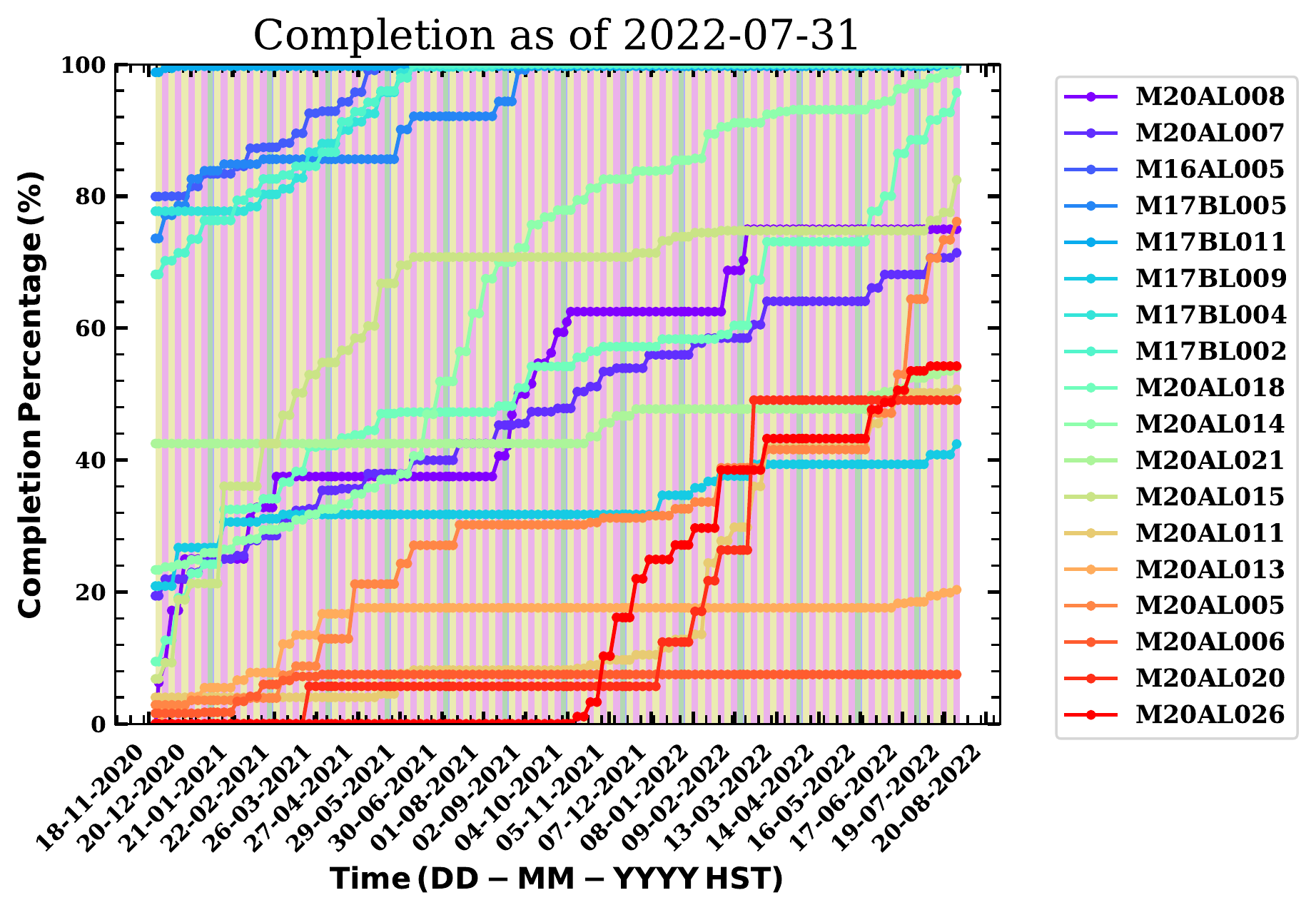}
\caption{Example program completion chart. The hours completed by each program (shown with different colors indicated on the legend) are tracked incrementally throughout the simulation. The JCMT queue structure (LAP: yellow, PI: magenta, DDT: green, UH: blue) is also indicated by the shaded regions, where observations are only performed during the LAP time blocks (with the exception of the ToO program M20AL008, which can run in LAP, PI or DDT blocks). Tracking the incremental progress of Large Programs can be useful in pinpointing periods where certain programs have stalled in their execution, and in turn, identifying the root cause of the problem.
}
\label{fig:prog_completion}
\end{figure}

\begin{figure}[t!]
\centering
\includegraphics[width=0.48\textwidth]{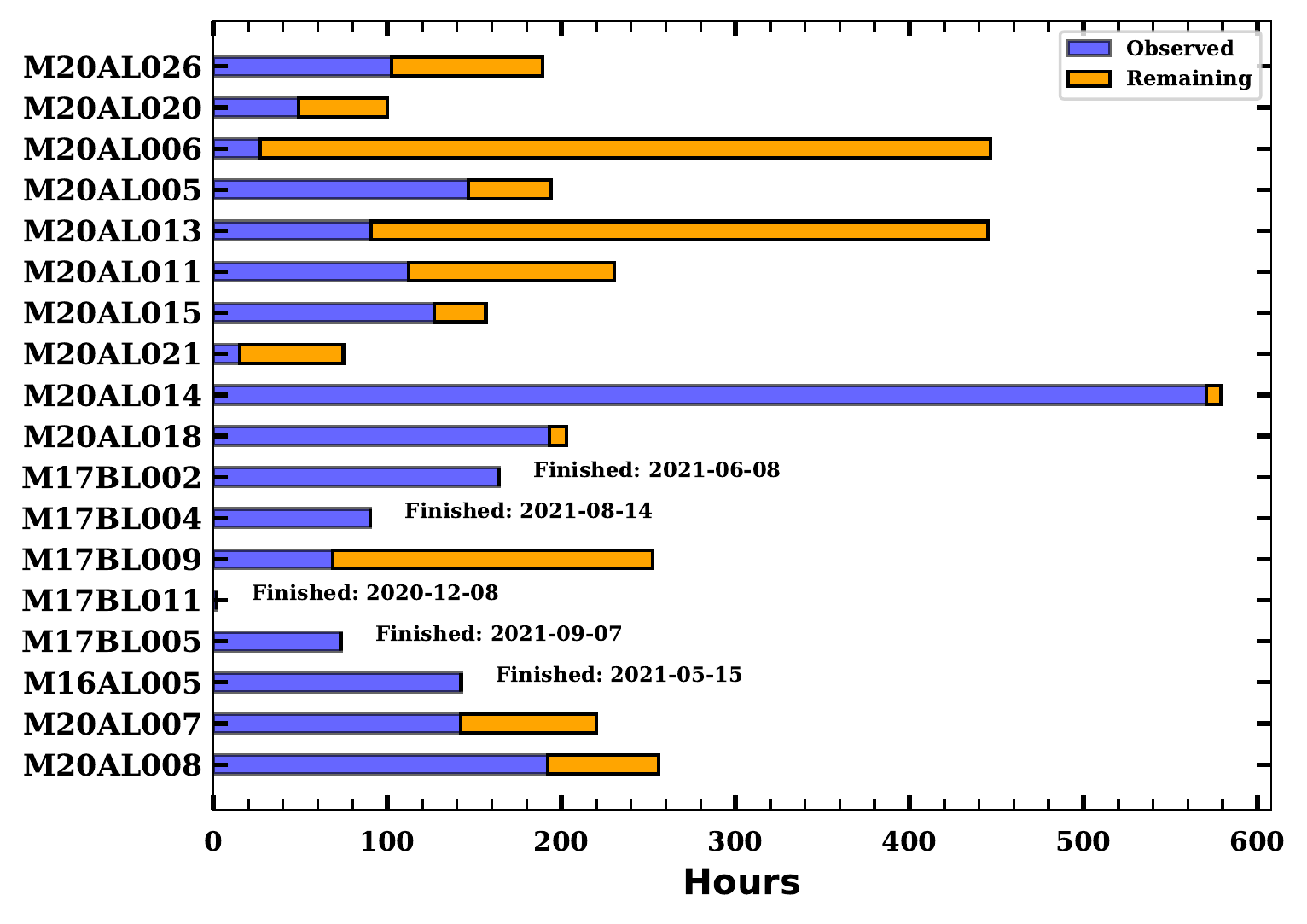}
\includegraphics[width=0.48\textwidth]{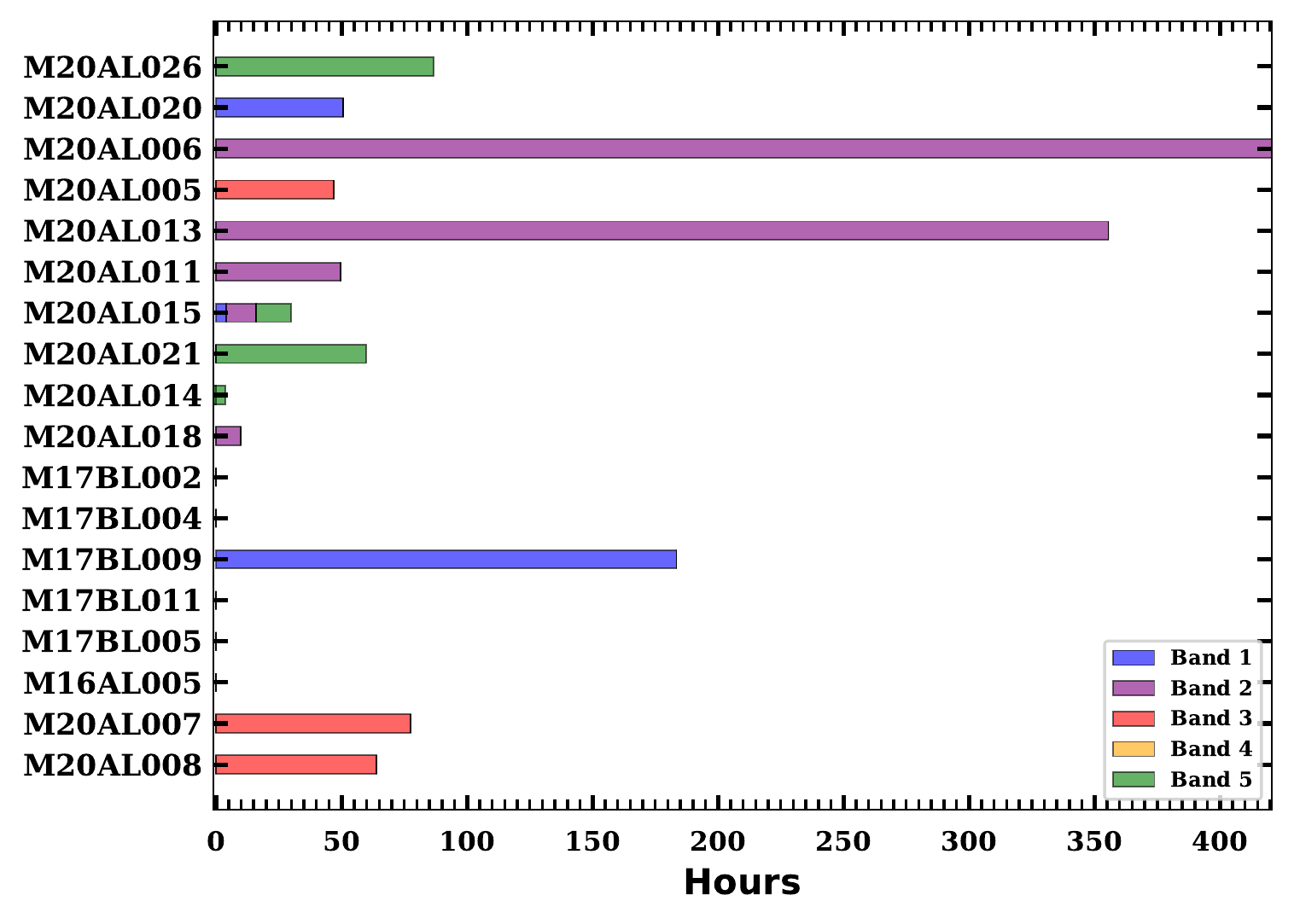}
\caption{Example bar charts displaying the overall results of a simulation run. {\it Left:} The hours completed (blue) and remaining (orange) are shown as fractions of the total allocated time for each program. If a program has finished during the simulation, the date at which the program finished is shown. {\it Right:} The remaining time after the simulation run split by program (shown by the different bars) and weather grade (shown by different colors indicated in the legend on the bottom right). Visualizing Large Program completion and the current weather grade requirements of each Large Program can be especially useful in determining when it is appropriate to offer a new call for more Large Programs at the telescope, and specifying the observing parameters for future Calls for Proposals at the JCMT. 
}
\label{fig:prog_res}
\end{figure}

\begin{figure}[t!]
\centering
\includegraphics[width=0.7\textwidth]{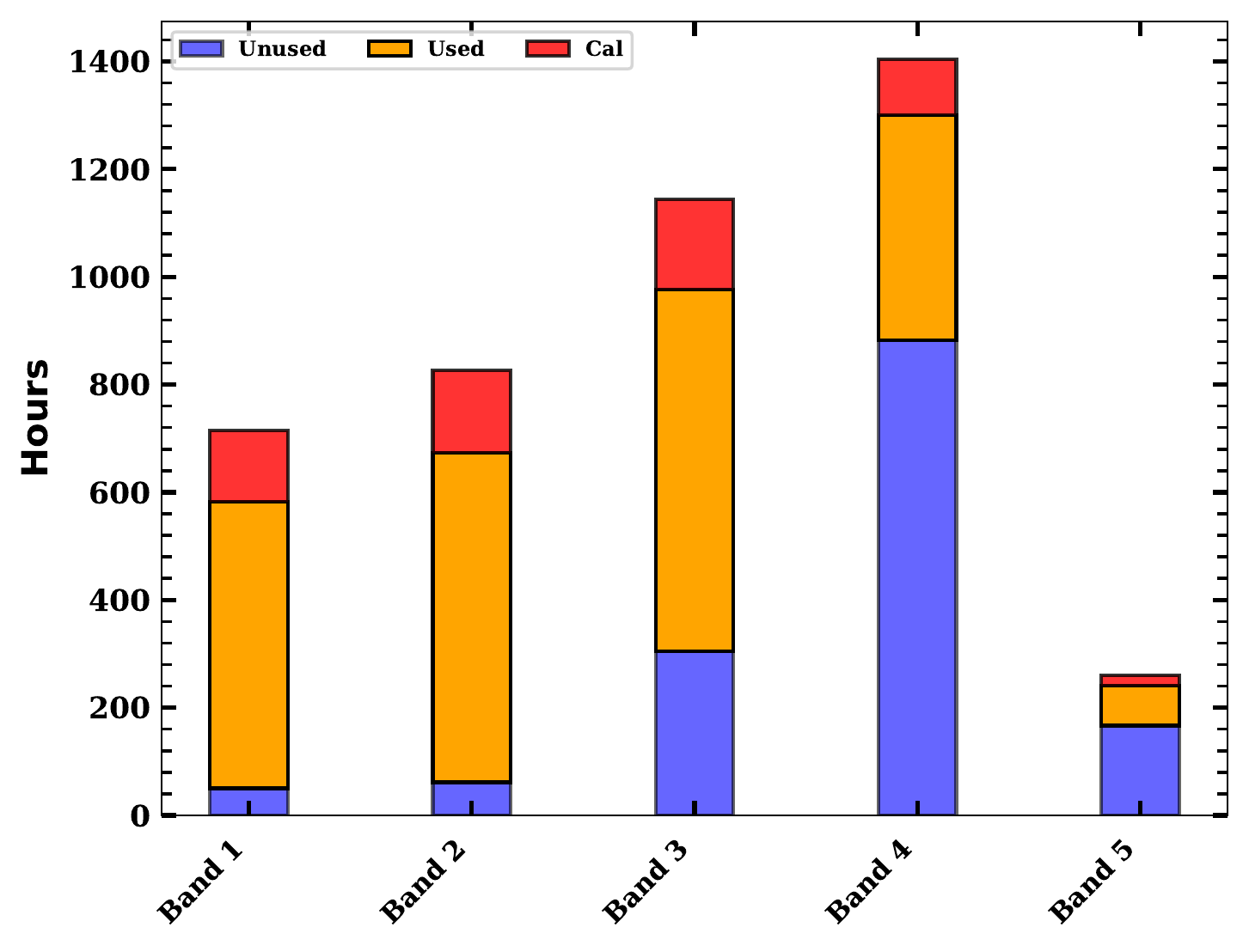}
\caption{Example bar chart displaying a weather grade time breakdown for a simulation run. The total hours available in each weather grade are shown, split into hours used by Large Programs (orange), hours unused where nothing was observed (blue), and hours on calibration sources (red). This statistic can be used to identify weather grades with a lack of Large Program targets, and in turn be informative in specifying the observing parameters for future Calls for Proposals at the JCMT.
}
\label{fig:unused_tally}
\end{figure}

\begin{figure}[t!]
\centering
\includegraphics[width=0.65\textwidth]{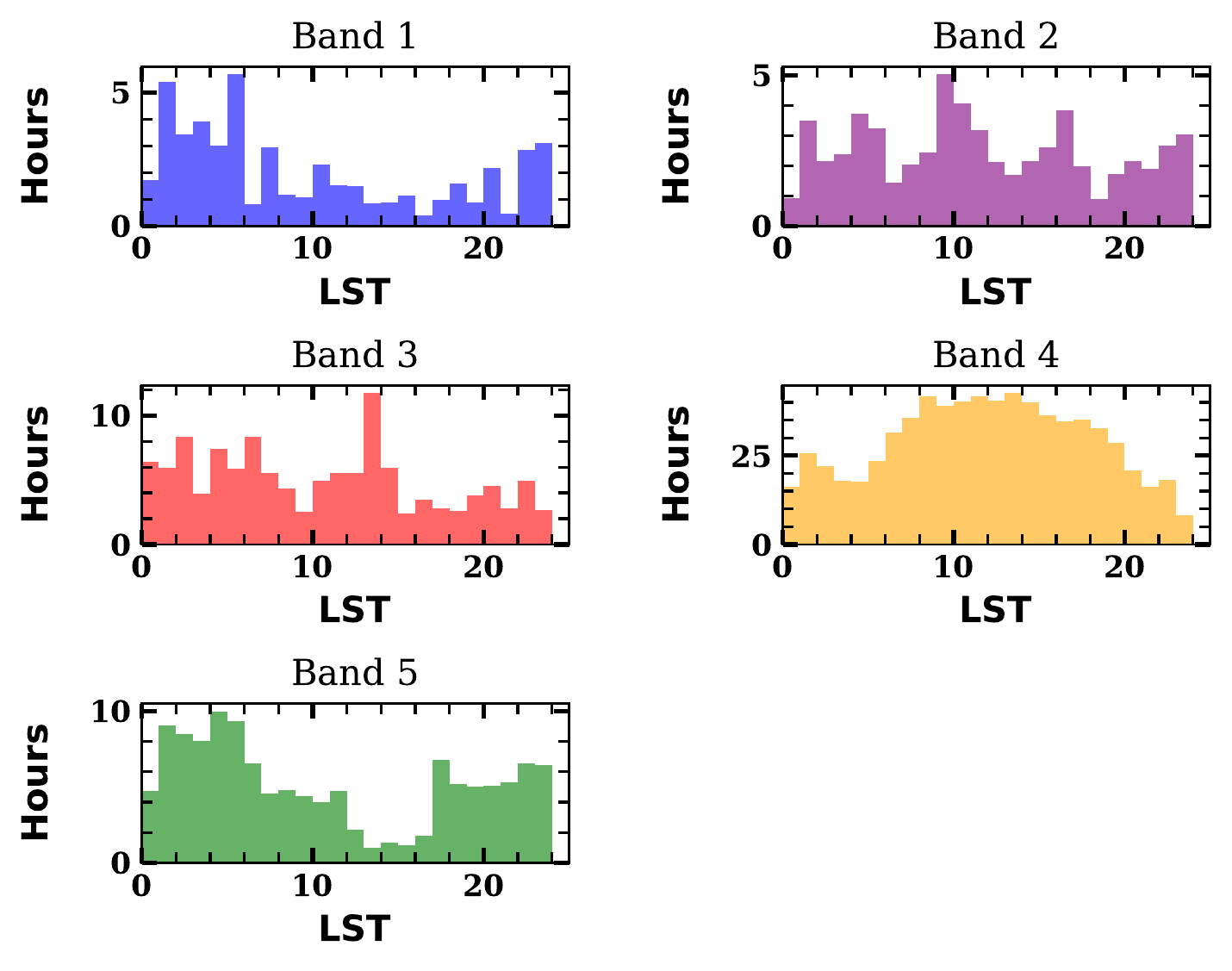}
\caption{Example histograms of the Local Sidereal Time (LST) blocks of unused time in each weather grade for a simulation run. One hour bins are used in these histograms. This statistic can be used to identify LST ranges that are not being filled by Large Program targets at the JCMT.
}
\label{fig:unused_ra_hist}
\end{figure}

\begin{figure}[t!]
\centering
\includegraphics[width=0.7\textwidth]{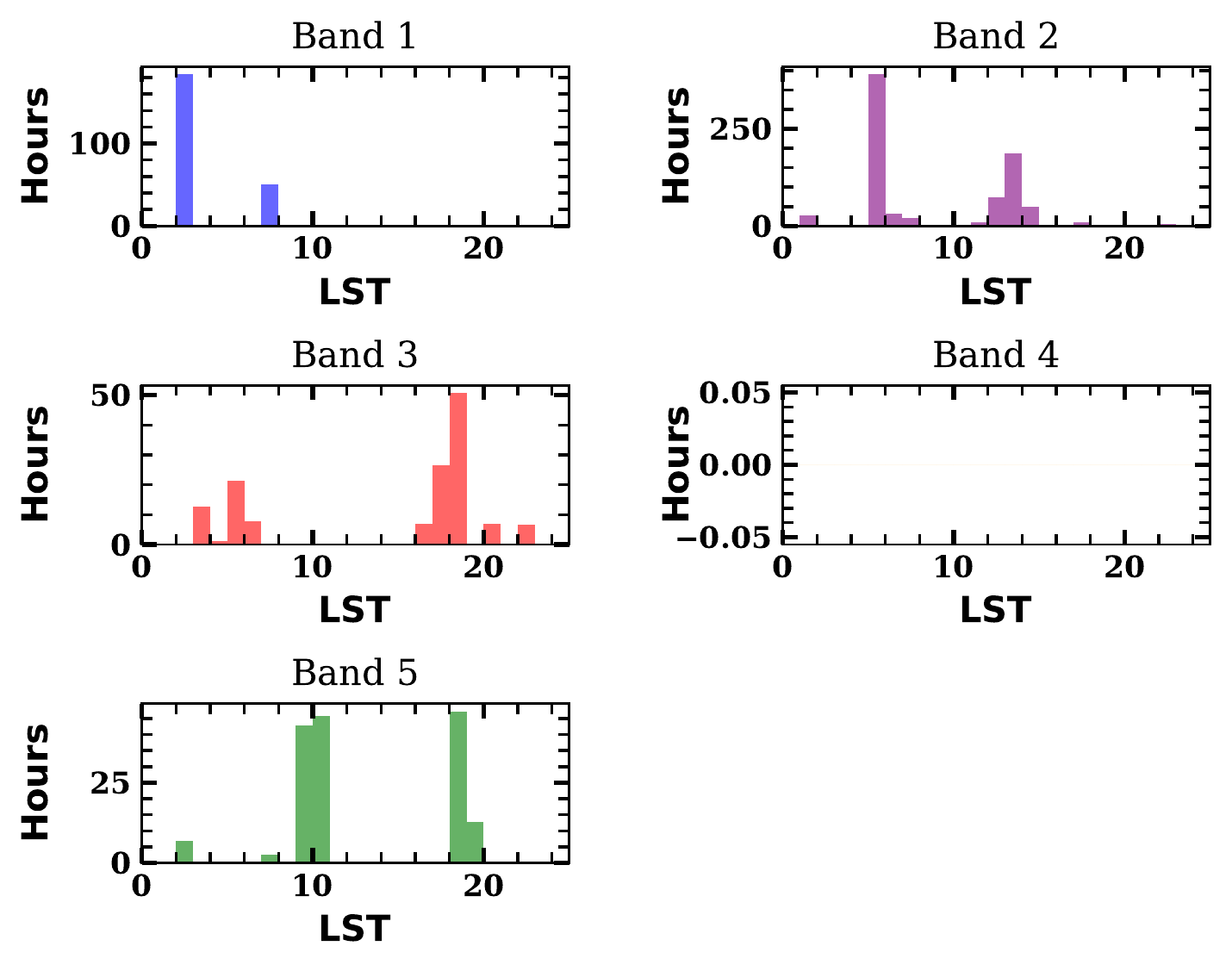}
\caption{Example histograms of the Local Sidereal Time (LST) blocks of remaining time for all of the Large Programs in each weather grade for a simulation run. One hour bins are used in these histograms. This statistic can be used to easily pinpoint the LST ranges of remaining Large Program targets yet to be observed.
}
\label{fig:remaining_ra_hist}
\end{figure}

\begin{figure}[t!]
\centering
\includegraphics[width=0.7\textwidth]{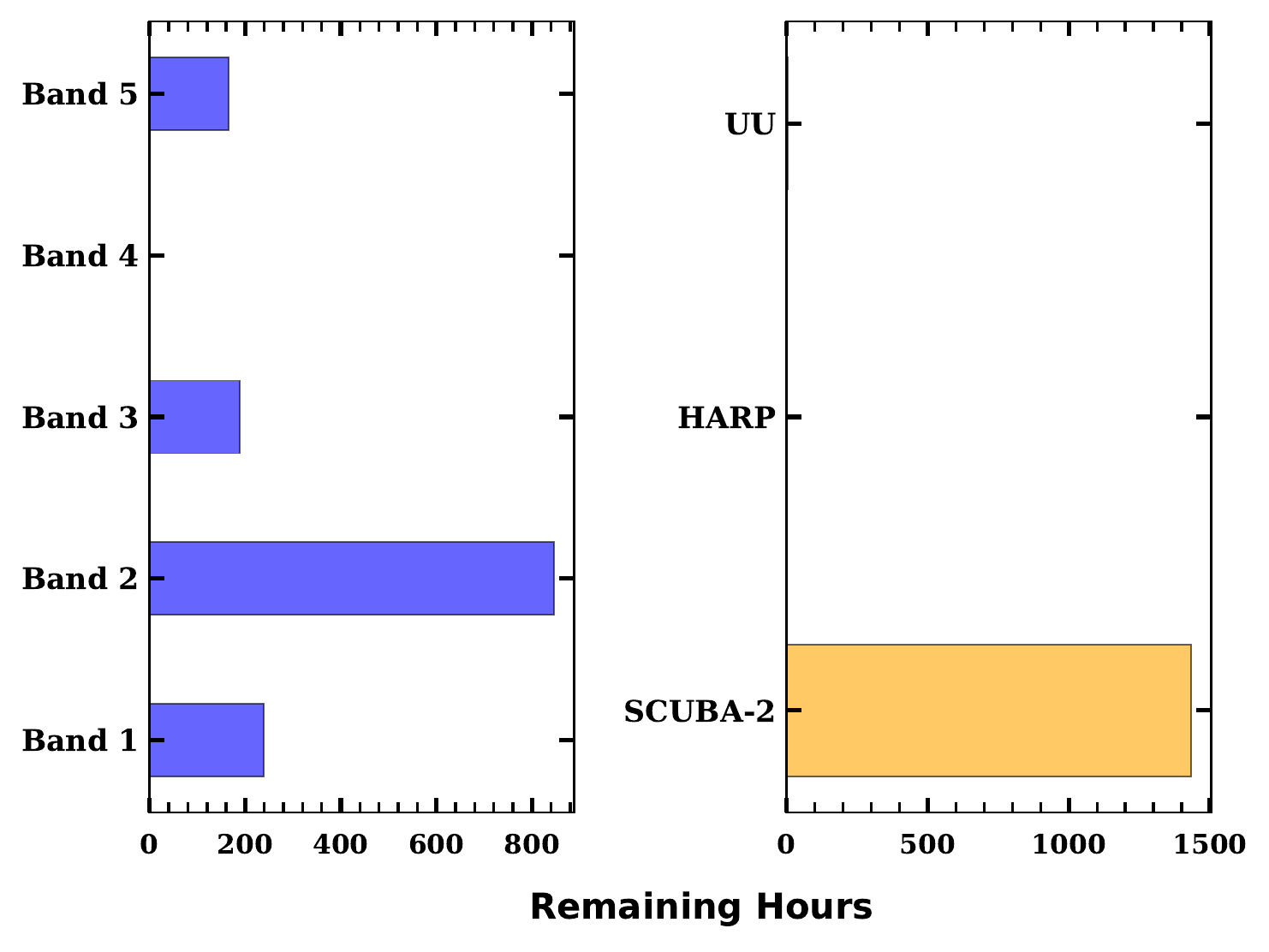}
\caption{Example bar chart displaying the remaining time in all the active Large Programs after a simulation run. Here the total remaining time is split by weather grade (\textit{left}) and by instrument (\textit{right}). These statistics can be used to identify which weather grades or instruments are currently being over/under-utilized by the current Large Programs.}
\label{fig:wplusinst}
\end{figure}

The {\sl JCMT Automated Project Completion Forecaster} produces four different text based files, and eight different figures to display the results of the simulation. The text based files include: (1) overall results (tabulating simulation hours and remaining hours for each program, program completion dates, and Large Program statistics, such as, total Large Program hours observed/remaining after the simulation and total hours lost to weather in the simulation), (2) a breakdown of available, used, and unused hours in each weather grade during the simulation, (3) the remaining time in the Large Program queue, split by weather grade, instrument, and program, and (4) tabulated schedules for each night in the simulation. Here we provide examples of the figures produced by the simulation; nightly schedule (Figure~\ref{fig:ex_sched_201917}), program completion chart (Figure~\ref{fig:prog_completion}), program results charts (Figure~\ref{fig:prog_res}), time breakdown chart (Figure~\ref{fig:unused_tally}), unused Local Sidereal Time (LST; equivalent to the right ascension on the observers meridian) histograms per weather grade (Figure~\ref{fig:unused_ra_hist}), LST histograms per weather grade for remaining Large Program targets after the simulation (Figure~\ref{fig:remaining_ra_hist}), and remaining hours for all Large Programs split by weather grade and instrument (Figure~\ref{fig:wplusinst}). All of these outputs are designed to provide useful visualizations of the data over different regions of the complex observing parameter space, and statistics to aide in the planning of future JCMT observing needs.


\subsection{Future Improvements}
\label{sec:future}
As this is an ongoing project, we are planning to implement improvements into the {\sl JCMT Automated Project Completion Forecaster} code-base, including folding in lessons learned from future testing of the forecaster. Additionally, we would like to improve upon how the forecaster handles calibration observations. Rather than just assume a fixed amount of time (currently 25\% of the science target observing time is added to inflate the observing blocks) is spent on calibrations, we propose a procedure where we would first create a test schedule to estimate how many targets are observed with each instrument, and then re-run the scheduler with the appropriate number of calibrators added to nightly target list. Ultimately, we aim to provide the observatory and users with more accurate information for future Calls for Proposals, scheduling, and execution of Large Programs at the JCMT, however, we may also work towards generalizing the code-base so that it can be used by other facilities.

\clearpage
%\appendix    %>>>> this command starts appendixes

\acknowledgments % equivalent to \section*{ACKNOWLEDGMENTS}       
The authors wish to extend a sincere thanks to Doug Johnstone, who made the push for the Legacy Surveys tracking and forecasting plan under the Joint Astronomy Center, and Holly Thomas, for setting up the original tracking and forecasting tools used on the Legacy Surveys.
The James Clerk Maxwell Telescope (JCMT) is operated by the East Asian Observatory (EAO) on behalf of The National Astronomical Observatory of Japan; Academia Sinica Institute of Astronomy and Astrophysics (ASIAA); the Korea Astronomy and Space Science Institute; Center for Astronomical Mega-Science (as well as the National Key R\&D Program of China with No. 2017YFA0402700). Additional funding support is provided by the Science and Technology Facilities Council of the United Kingdom and participating universities and organizations in the United Kingdom and Canada. Additional funds for the construction of SCUBA-2 were provided by the Canada Foundation for Innovation. N\={a}makanui was constructed and funded by ASIAA in Taiwan, with funding for the mixers provided by ASIAA, and at 230 GHz by EAO. The N\={a}makanui instrument is a backup receiver for the Greenland Telescope (GLT). 
The authors also wish to recognize and acknowledge the very significant cultural role and reverence that the summit of Maunakea has always had within the indigenous Hawaiian community.  We are most fortunate to have the opportunity to conduct observations from this mountain.

% References
\bibliography{report} % bibliography data in report.bib
\bibliographystyle{spiebib} % makes bibtex use spiebib.bst

\end{document}